# Chirality-Induced Orbital-Angular-Momentum Selectivity in Electron Transmission and Scattering

Yun Chen[1], Oded Hod[1], Joel Gersten[2], and Abraham Nitzan[1,3]

[1]*Department of Physical Chemistry, School of Chemistry, The Raymond and Beverly Sackler Faculty of Exact Sciences and The Sackler Center for Computational Molecular and Materials Science, Tel Aviv University, Tel Aviv 6997801, Israel*

[2]*Department of Physics, City College of the City University of New York, New York, New York 10031, USA*

[3]*Department of Chemistry, University of Pennsylvania, 231 South 34th Street, Philadelphia, Pennsylvania 19104, USA*

## Abstract

Chirality-induced orbital-angular-momentum selectivity (CIOAMS) in electron transmission and scattering processes is investigated. Polarization of the OAM of an electron traversing chiral media is first studied via electronic wavepacket propagation using the time-dependent Schrödinger equation. Next, spatial resolution of wavepackets carrying opposite OAM, following scattering from a corrugated surface is demonstrated. This suggests that OAM may play a significant role in the mechanisms underlying chirality induced spin selectivity, measured for electrons crossing chiral media in setups involving Mott polarimetry. Our results highlight the potential to exploit CIOAMS in innovative emerging quantum technologies.

## Introduction

Chirality arises when parity symmetry is absent or disrupted, resulting in non-superimposable mirror images of an object. This property is ubiquitous in various natural phenomena and holds profound significance in disciplines such as chemistry, biology, medicine, physics, and astronomy[1-5]. When electrons traverse chiral media, such as helical molecules or those with chiral centers, their transmission probability is found to depend on the orientation of their spin vector relative to the propagation direction. This effect, termed chirality-induced spin selectivity (CISS)[6-17], reveals a fundamental interplay between the spin degree of freedom of subatomic particles and their chiral molecular environment, and paves the way for groundbreaking technologies in spintronics[6, 8, 12, 18] and enantioseparation[19-24].

Since its discovery in photoemission experiments[25], the CISS effect has been observed for numerous chiral materials and in various experimental manifestations[17]. Nonetheless, despite extensive efforts invested in demonstrating and rationalizing the CISS effect, the underlying microscopic origins and physical mechanisms remain elusive. Naturally, the relation between electron spin and molecular chirality must rely on spin-orbit coupling (SOC). However, the relatively small SO interactions characterizing hydrocarbon-based molecules, in which significant CISS filtering has been observed, cannot provide a quantitative explanation for the experimental findings, especially at room temperature[14]. In response to this challenge, several theoretical frameworks have been proposed to account for the apparent amplification of SOC effects in chiral organic molecules, aiming to bridge the gap between theoretical predictions and experimental observations. To that end, various physical ingredients have been considered, including electron correlations[26, 27], molecular vibrations[28, 29], energy dissipation[30-33], non-unitary dynamics[34], and spinterface effects[35-37].

Additional mechanisms for CISS that have been proposed involve orbital angular momentum (OAM) selection within a chiral medium overlying a substrate of strong SOC[36, 38-44]. Gersten et al.[38] introduced the concept of induced spin filtering, where a chiral medium of low SOC filters OAM of the traversing electrons, which in turn correlates with the spin angular momentum due to strong SOC in the substrate from which the electron arrived. A complementary idea was suggested by Liu et al.[40], where the weak-SOC chiral medium is thought to polarize electron OAM via orbital-momentum locking, which is then converted to spin polarization in the strong-SOC outgoing electrode. These theories recently gained experimental support using magnetic semiconductor-based chiral molecular spin valves coupled to electrodes of different SOC[45]. Nevertheless, the underlying mechanisms fail to explain spin-dependent photoemission experiments through chiral molecular systems residing atop Si[46], Cu[47] and Al[48] substrates of weak SOC, as well as CISS observations in other processes that do not involve metallic substrates[49].

One of the central experimental tools to investigate the origins of the CISS effect in electron emission setups is Mott polarimetry[6, 50, 51]. Here, electrons that cross a chiral medium are accelerated and then scattered off a heavy atom surface of high spin-orbit coupling. This leads to an angular distribution of the scattered electrons that is interpreted as the outcome of spin polarization. In light of the proposed CISS mechanisms that involve OAM screening within the chiral medium, a question arises whether the scattering process may also involve OAM-based resolution.

A simple picture exemplifying the idea is that of a spinning classical ball scattering off a surface. If the surface is smooth and frictionless then the recoil direction of the ball is independent of its initial angular momentum about its center of mass (COM). Nonetheless, once friction is introduced, clockwise and counterclockwise spinning balls with angular momentum parallel to the surface will scatter in different trajectories, resulting in angular-momentum-based spatial resolution. This effect was recently used to demonstrate a classical spin-off of the CISS effect, where molecules carrying opposite angular momenta were shown to be spatially resolved when traversing frictional helical channels[52]. Alternatively, if surface roughness is introduced and the ball is flexible, its collision dynamics and hence its scattering direction will depend on its spinning sense, even in the absence of surface friction, manifesting energy dissipation processes to internal degrees of freedom of the ball. While quantum particles are not expected to scatter like classical bodies[53], one may expect that dissipative or rough surfaces will scatter electronic wavepackets that carry opposite OAM in different quantum trajectories as well.

To investigate this hypothesis, we performed single particle wavepacket simulations of the two central processes involved in electron emission CISS experiments: (i) OAM polarization of quantum electronic wavepackets traversing chiral media, and (ii) scattering of electronic wavepackets that carry well-defined OAM from different surfaces. These simulations demonstrate that passage through a helical potential field leads to buildup of orbital angular momentum of a traversing electron and that the spatial angular distribution of an electron that is scattered from a corrugated surface is significantly affected by (and therefore carries distinct information about) the orbital angular momentum of the incident electronic wavefunction.

## Results and discussion

Our model system for the chirality-induced OAM polarization simulations consists of an electronic Gaussian wavepacket of the following initial form:

$$\psi(x,y,z;t=0) = \frac{1}{\sqrt{\sqrt{\pi}^3 \sigma_x \sigma_y \sigma_z}} e^{-\frac{(x-x_0)^2}{2\sigma_x^2}} e^{-\frac{(y-y_0)^2}{2\sigma_y^2}} e^{-\frac{(z-z_0)^2}{2\sigma_z^2}} e^{i\mathbf{k}^0 \cdot \mathbf{r}}, \qquad [1]$$

where $\mathbf{r} = (x, y, z)$ is the position vector, $\sigma_x$, $\sigma_y$, and $\sigma_z$ represent the initial widths of the wavepacket in the $x$, $y$ and $z$ directions, $\mathbf{r}_0 = (x_0, y_0, z_0)$ denotes the initial position of its COM, and $\mathbf{k}^0 = (k_x, k_y, k_z)$ is the initial wavevector that sets its group velocity. The wavepacket is driven through a rigid helix of positively charged point particles by a uniform static electric field (see Fig. 1a) and the expectation values of its position and angular momentum are recorded. Further details regarding the model system and the simulation setup are provided in the Methods section.

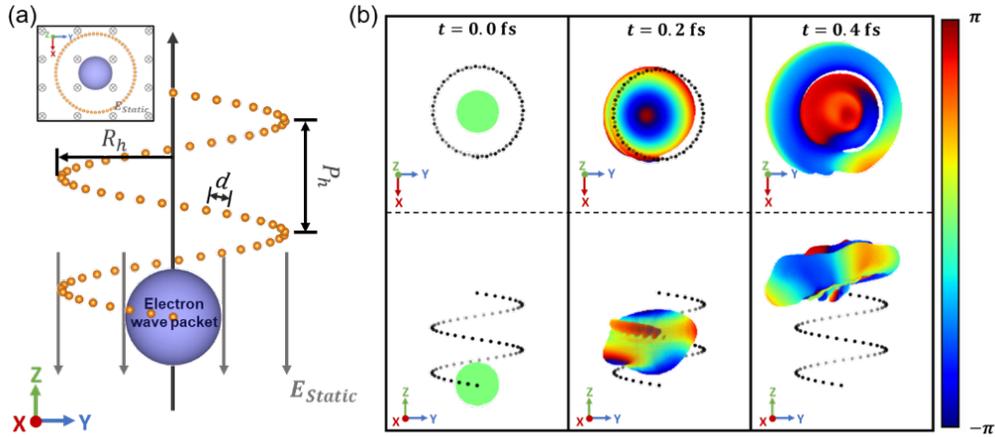

**Figure 1**. (a) Schematic illustration of the model system used for the chirality-induced OAM polarization simulations. The model consists of an initial Gaussian quantum wavepacket (purple sphere) driven by a uniform electric field through a rigid left-handed helix of positively charged fixed point particles. (b) Snapshots of the wavepacket evolution and its spatial phase distribution (see colorbar on the right) at $t = 0, 0.2, and\ 0.4\ fs$ (both top and side views of each snapshot are provided in the upper and lower panels, respectively). The initial width of the wavepacket is $\sigma_x = \sigma_y = \sigma_z = 1$ Å, the helix radius is $R_h = 5$ Å, its pitch length is $P_h = 5$ Å with a total of two pitches, along which the fixed particles of charge $q = 1\ a.u.$ are uniformly spread with an interparticle arclength spacing of $d = 1$ Å. The wavepacket is given no initial momentum and is driven upwards by an electric field of $E = (0,0,-5)\ V/$Å.

Figure 1b presents snapshots along the wavepacket trajectory, indicating that as it ascends the helix under the external field the initial spherical packet accumulates angular momentum due to its attraction to the charged helical backbone until it exits the top of the helix in a swirling mushroom-like structure (see Supporting Information (SI) Movie 1), showing a shape reminiscent of a vortex electron wavefunction[54, 55]. This demonstrates that electronic linear momentum can be converted into angular momentum of well-defined rotational sense through the torque exerted by a chiral field.

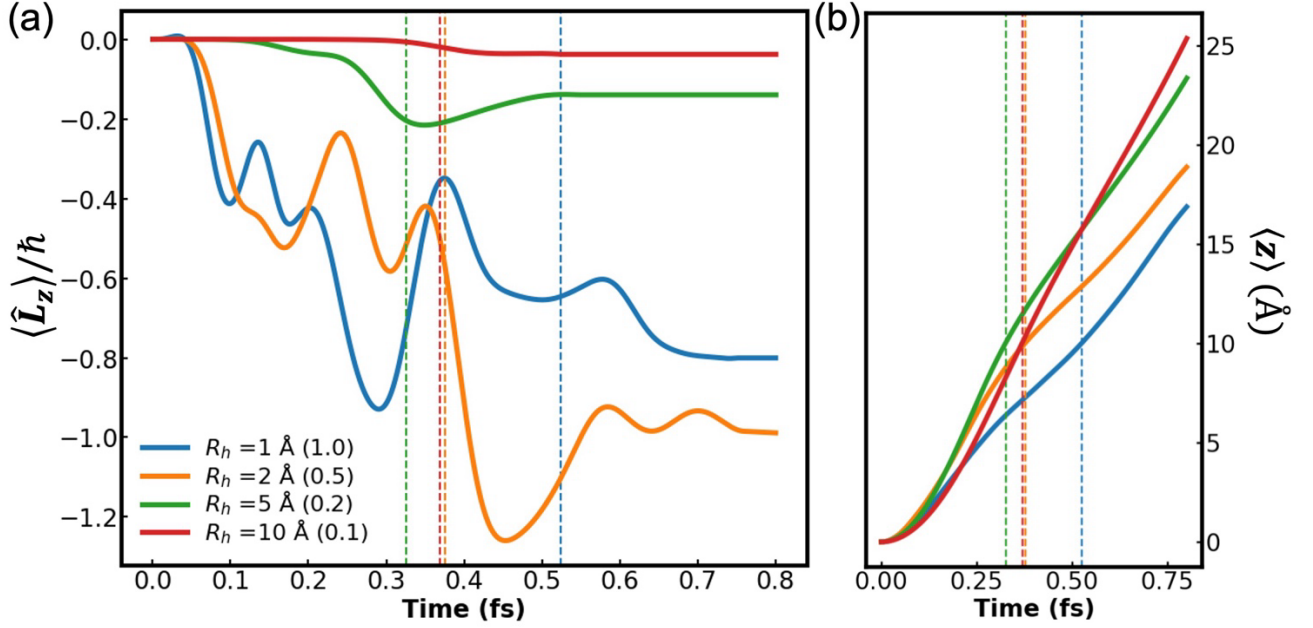

**Figure 2.** (a) Time evolution of the $\langle \hat{L}_z \rangle(t)$ angular momentum expectation value component of an initial Gaussian wavepacket of $\sigma_x = \sigma_y = \sigma_z = 1$ Å traversing left-handed helixes of $R_h = 1$ (blue line), 2 (orange line), 5 (green line), and 10 (red line) Å (all other system parameters are the same as those provided in the caption of Fig. 1). The ratios between the initial wavepacket width and the different helix radii, $\tau = \sigma/R_h$, are given in parentheses. (b) Time evolution of the vertical position expectation value component of the wavepacket, $\langle \hat{z} \rangle(t)$, where the same color code is used as in panel a. The dashed lines in both panels indicate the time at which $\langle \hat{z} \rangle(t)$ crosses the top of the helix.

To evaluate the dependence of the efficiency of linear to angular momentum conversion on system parameters, we performed a set of comparative simulations. First, we investigated the impact of the ratio between the initial wavepacket dimensions and the helix radius on angular momentum accumulation. To that end, we considered a spherical Gaussian wavepacket of initial width of $\sigma_x = \sigma_y = \sigma_z = 1$ Å passing through charged left-handed helices of radii $R_h = 1, 2, 5,$ and $10$ Å (all other simulation parameters are the same as those provided in the caption of Fig. 1). Fig. 2(a) presents the time evolution of the OAM expectation value component along the main axis of the helix, $\langle \hat{L}_z \rangle(t)$. Clearly, when the lateral dimensions of the helix are comparable to the width of the wavepacket (blue and orange lines), the latter accumulates considerable angular momentum, whereas for wider helices (green and red lines) the weaker attraction of the wavepacket to the charged helix results in reduced angular momentum gain rate.

As long as the wavepacket resides within the helix, its attraction to the top and bottom sections of the helix is balanced and the vertical motion of its center of charge is mainly dictated by the external field. However, when the wavepacket exists at the top of the helix it experiences a downwards attraction to the charged helix, opposing the external field, that results in a reduction of the vertical velocity (see slope reduction at the time window of $0.25 - 0.5$ fs in Fig. 2b). This pulling down of the wavepacket back towards the helix is accompanied also by a torque in the opposite rotational sense due to the chiral field. This, in turn, results in a reduction of the accumulated angular momentum (see orange curve in Fig. 2a at ~0.45 fs), which eventually stabilizes once the wavepacket is sufficiently far from the helix. Further details on the dependence of orbital polarization on helix radius (for fixed initial wavepacket dimensions) are provided in SI Section 1.

The helix pitch is another parameter that may strongly affect angular momentum buildup. This is demonstrated in Fig. 3, which compares the vertical angular momentum expectation value, $\langle \hat{L}_z \rangle(t)$, accumulated by the wavepacket as it passes through helices of $P_h = 5$ (blue line), 7.5 (orange line), and 10 (green line) Å. The results reveal that helices of longer pitch (within the range considered) exhibit higher angular momentum accumulation. This is attributed to the longer interaction time that the wavepacket experiences with the charged helical chain. Naturally, should the pitch length significantly exceed the initial wavepacket dimensions $(P_h \gg \sigma_x = \sigma_y = \sigma_z)$ the angular momentum accumulation would be negligible, suggesting that there is an optimal ratio between the helix pitch and the initial wavepacket width at which angular momentum accumulation is maximal. These results demonstrate that by controlling the geometric structure of the chiral medium (via, e.g., its chemical composition) one can dictate the spatial angular momentum accumulated by traversing electrons.

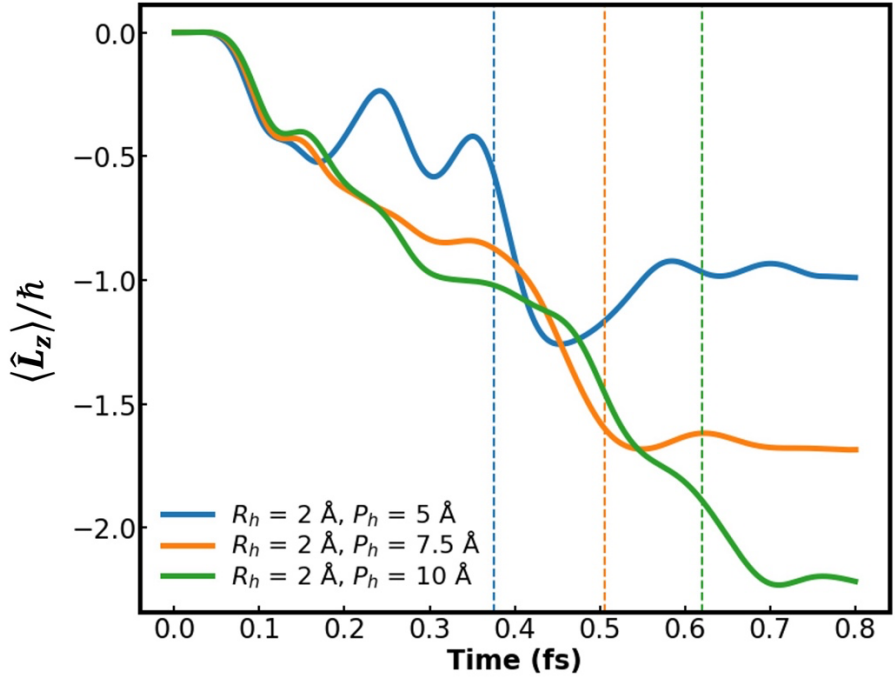

**Figure 3**. Time evolution of the $\langle \hat{L}_z \rangle(t)$ angular momentum expectation value component of an initial Gaussian wavepacket of $\sigma_x = \sigma_y = \sigma_z = 1$ Å traversing left-handed helixes of $P_h = 5$ (blue line), 7.5 (orange line), and 10 (green line) Å with $R_h = 2$ Å (all other system parameters are the same as those provided in the caption of Fig. 1). The dashed lines indicate the time at which $\langle \hat{z} \rangle(t)$ crosses the top of the helix.

Having established that chiral environments indeed induce electronic OAM polarization on traversing electrons, we now turn to investigate the scattering of such vortex electrons from surfaces. These simulations are carried out in two dimensions (2D, see Fig. 4a). The initial incident electron wavefunction is taken to be:

$$\psi(x, y; t = 0) = A\left(\sqrt{(x-x_0)^2 + (y-y_0)^2}\right)^{|m|} e^{im\phi} e^{-\frac{(x-x_0)^2}{2\sigma_x^2}} e^{-\frac{(y-y_0)^2}{2\sigma_y^2}} e^{ik_y y}, \qquad [2]$$

which describes a normalized ($A$ being the normalization factor) 2D Gaussian wavepacket located at $r_0 = (x_0, y_0)$, carrying OAM of $\hbar m$ about an axis pointing in $z$ direction and crossing the center of the wavepacket ($\phi$ being the azimuthal angle about this axis), and given momentum of $\hbar k_y$ in the y direction (see Fig. 4a)[56]. The wavepacket propagates towards, and then scatters from, a laterally smooth (in the $x$-direction) Lennard-Jones (LJ) wall of the form:

$$V_{LJ}(y) = 2\varepsilon_{LJ}\left[\frac{1}{2}\left(\frac{y_{min}}{y}\right)^{12} - \left(\frac{y_{min}}{y}\right)^{6}\right] \qquad [3]$$

with $\varepsilon_{LJ} = 0.05$ eV, which lies along the $y$ axis with its minimum located at $y_{min} = -2.25$ Å from the top edge of the simulation box (defined as the vertical origin).

Similar to classical dissipationless scattering, the spinning wavepacket backscatters vertically from the wall, regardless of its initial OAM. Surprisingly, contrary to classical rigid ball scattering, the collision process is found to not only invert the linear momentum of the quantum wavepacket but also its angular momentum (see Fig. 4b). An initial wavepacket spinning counterclockwise about its COM with $m = 1$, backscatters into a clockwise spinning state with $m = -1$, and vice versa. The difference arises from the fact that while the classical rigid ball hits the wall with a single (or very localized) contact point, it is the entire quantum wavepacket that interacts with the wall potential. Focusing, for example, on two opposite density pixels on the equatorial line of a spinning Gaussian wavepacket that propagates towards the wall (see Fig. 4a, middle), the density at one cell has a local vertical velocity of $v_y^1 = v_y^{COM} + \omega r$, whereas for the other cell it is $v_y^2 = v_y^{COM} - \omega r$. When these two density cells collide with the wall their vertical velocity flips, yielding $v_y^1 = -v_y^{COM} - \omega r$ and $v_y^2 = -v_y^{COM} + \omega r$, thus resulting in reversal of the spinning sense of the wavepacket (see Fig. 4a, right). Mathematically, this can be rationalized by considering the time evolution of the expectation value of the perpendicular angular momentum operator component, $\langle \hat{L}_z \rangle$:

$$\frac{d\langle \hat{L}_z \rangle}{dt} = \frac{d\langle \psi | \hat{L}_z | \psi \rangle}{dt} = \frac{d\langle \psi |}{dt} \hat{L}_z | \psi \rangle + \langle \psi | \hat{L}_z \frac{d | \psi \rangle}{dt} = \frac{i}{\hbar} \langle \psi | \hat{H} \hat{L}_z | \psi \rangle - \frac{i}{\hbar} \langle \psi | \hat{L}_z \hat{H} | \psi \rangle = \frac{i}{\hbar} \langle \psi | [\hat{T} + \hat{V}, \hat{L}_z] | \psi \rangle, \quad [4]$$

where we used the time-dependent Schrödinger equation. Because the kinetic energy operator, $\hat{T}$, commutes with $\hat{L}_z$, Eq. [4] demonstrates that variation of the perpendicular angular momentum expectation value during collision requires that $\hat{L}_z$ and the potential operator are non-commutative. This, indeed, is the case in our simulations, where:

$$[V_{LJ}(y), \hat{L}_z] = [V_{LJ}(y), x\hat{P}_y - y\hat{P}_x] = i\hbar x \frac{\partial V_{LJ}(y)}{\partial y} \neq \hat{0} \quad [5]$$

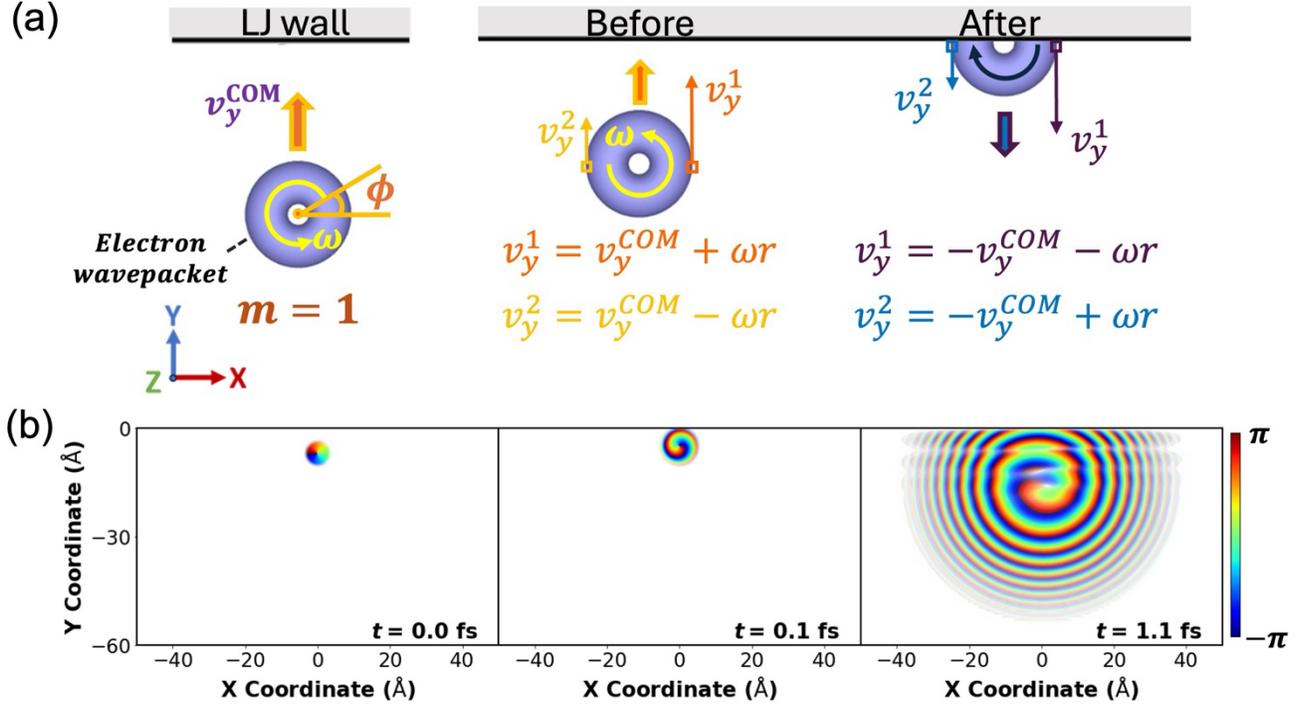

**Figure 4**. (a) Schematic illustration of the scattering of a 2D Gaussian wavepacket of angular momentum quantum number $m = 1$ from a horizontal LJ potential wall. Left: illustration of the spinning Gaussian wave packet approaching the wall (the angular coordinate $\phi$ is defined). Middle: illustration of the velocity of two opposite density pixels, located at a distance $r$ from the center of the wavepacket, along its equatorial line, before scattering. Right: illustration of the velocity inversion of the same density pixels after the collision, leading to reversal of the spinning sense of the wavepacket. (b) Phase-resolved snapshots extracted from the time-evolution of a wavepacket of an initial width of $\sigma_x = \sigma_y = 1$ Å, given an initial COM velocity of $v_y = 20$ Å/fs towards the wall, while carrying an initial angular momentum of $m = +1$, before ($t = 0$ and 0.1 fs) and during ($t = 1.1$ fs) collision. Similar results (with opposite rotational senses) are obtained for an initial $m = -1$ wavepacket (see SI movie 2).

The reversal of the angular momentum of the wavepacket does not affect its COM trajectory. Hence, it cannot be used on its own to achieve OAM spatial resolution. Nonetheless, it does demonstrate the unique asymmetric surface scattering behavior of a spinning wavepacket, which can be harnessed to manipulate wavepackets of opposing OAM to scatter along different trajectories. This can be achieved by introducing atomic surface corrugation and/or energy dissipation. To demonstrate the former, we first place an obstacle in the trajectory of the spinning wavepacket (see Fig. 5a and SI Section 2). The obstacle consists of a fixed point particle, located at $r_P$, that interacts with the wavepacket through a repulsive central potential for the form:

$$V_\text{P}(\boldsymbol{r}) = Ae^{-|\boldsymbol{r}-\boldsymbol{r}_\text{P}|/b}, \qquad [6]$$

where $A$ is the repulsion strength and $b$ is the interaction length. For head-on scattering of a wavepacket that does not spin about its center of charge ($m = 0$, middle row in Fig. 5b) diffraction evolves symmetrically around the particle. Once given angular momentum, the diffraction pattern becomes asymmetric, with $m = \pm 1$ resulting in mirror image trajectories (see top and bottom rows in Fig. 5b and SI Movie 3). For a counterclockwise spinning wavepacket of $m = 1$ the front pixels carry a left pointing lateral velocity component (towards the negative $x$ axis, see red arrow in the top middle subpanel of Fig. 5b) that upon reflection from the fixed scatterer results in a deflection of the wavepacket to the right. For the corresponding clockwise spinning $m = -1$ counterpart the opposite process occurs (see bottom subpanels of Fig. 5b). Fig. 5c presents the time evolution of the lateral expectation value component, $\langle x \rangle$, of the wavepacket for different initial angular momenta ($m = +1, 0,$ and $-1$) and various initial velocities. Regardless of the initial velocity, non-spinning wavepackets diffract without lateral deflection, whereas OAM carrying wavepackets that spin (counter)clockwise deflect to the (right) left, resulting in pronounced spatial resolution. Notably, a non-monotonic dependence of the transversal deflection on the incident velocity is obtained, where the $v_y = 30$ Å/fs (orange lines) case results in higher spatial separation than the $v_y = 10$ (blue lines) and $50$ Å/fs (green lines) cases. This can be attributed to two competing effects: as the incident velocity increases the distortion of the wavepacket due to its scattering from the obstacle grows but the effective collision time decreases (see inset of Fig. 5c). Accordingly, at constant incident velocity, the transversal deflection of the wavepacket during the scattering process should grow with increasing repulsive interaction. This is indeed the case, as can be seen in Fig. 5d, where higher values of the repulsion parameter, $A$, result in larger lateral motion of the wavepacket accompanied by stronger vertical deceleration (see inset of Fig. 5d).

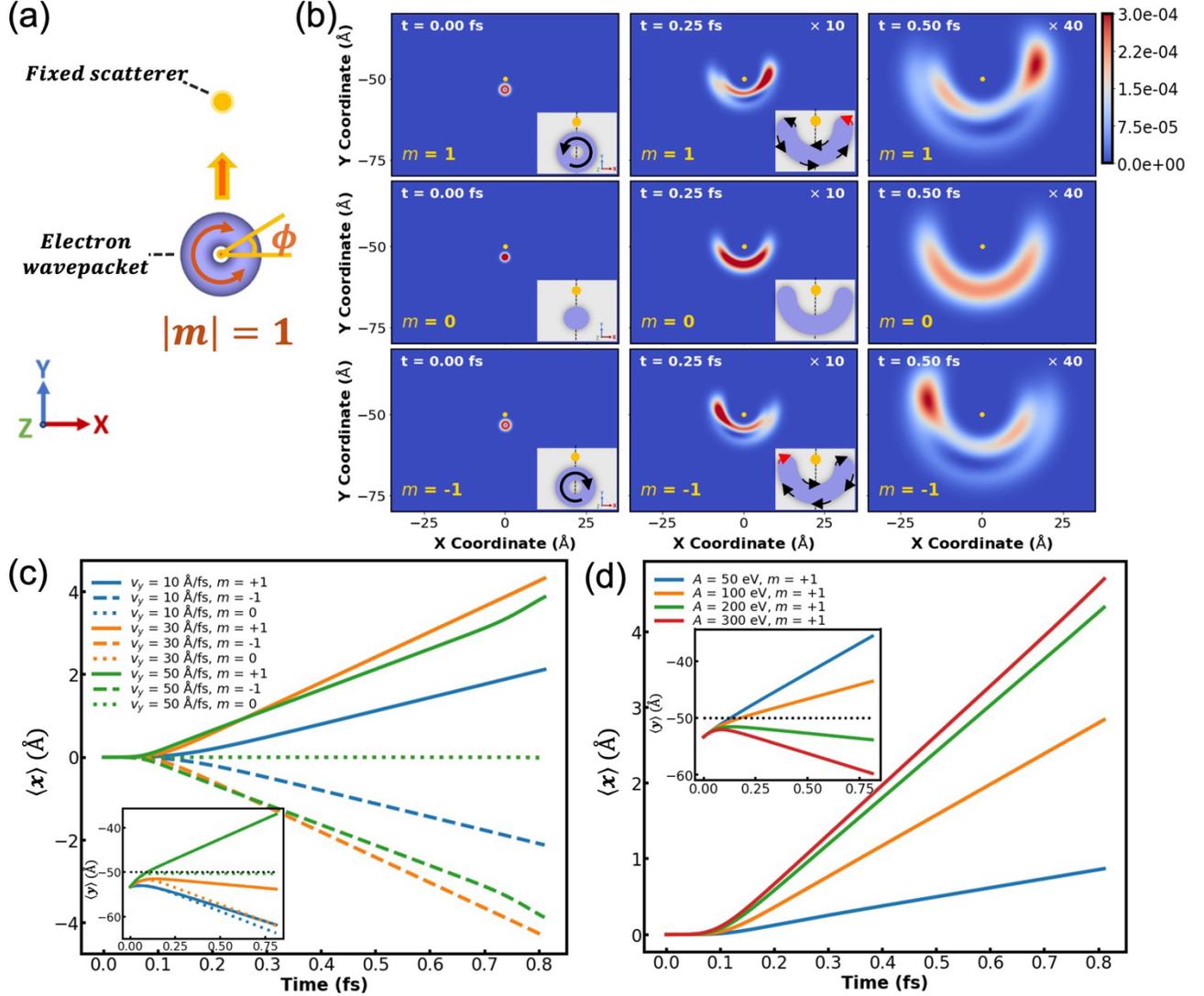

**Figure 5.** (a) Schematic illustration of the scattering of a spinning (angular momentum quantum number $|m| = 1$) 2D electronic wavepacket from a fixed repulsive central potential. (b) Snapshots of the scattering process at $t = 0$ (left panels), 0.25 (middle panels), and 0.5 (right panels) fs for wave packets carrying initial angular momentum of $m = 1$ (top panels), 0 (middle panels), and $-1$ (bottom panels). The insets illustrate the spinning senses of the corresponding wavepackets. The simulation parameters are $\sigma_x = \sigma_y = 1$ Å, $v_y = \hbar k_y / m_e = 30 \frac{\text{Å}}{\text{fs}}$, $A = 200$ eV, and $b = 1$ Å. (c) Time evolution of the transversal coordinate expectation value, $\langle \hat{x} \rangle$, of a scattering wavepacket carrying an initial OAM z-component of $m = -1$ (dashed lines), 0 (dotted lines), and $+1$ (solid lines) and given an initial velocity of $v_y = 10$ (blue lines), 30 (yellow lines) and 50 (green lines) Å/fs. (d) Time evolution of the transversal coordinate expectation value, $\langle \hat{x} \rangle$, of a scattering wavepacket carrying an initial OAM z-component of $m = +1$ and experiencing a repulsion strength of $A = 50$ (blue line), 100 (yellow line), 200 (green line) and 300 (red line) eV. All other simulation parameters are the same as in (b). The insets present the time evolution of the vertical coordinate expectation value, $\langle \hat{y} \rangle$, of the scattering wavepacket, where the dashed black line represents the vertical position of the center of the scattering potential.

These results suggest that replacing the laterally smooth LJ potential wall with a structured wall may lead to spatial resolution of scattering wavepackets that carry opposite OAM. To investigate this, we augment the LJ potential with a structured surface, modeled by two rows of fixed scatterers, each similar to the one discussed above (see Fig. 6a and SI Section 3). The two rows are laterally shifted with respect to each other by half the lattice vector $(d/2)$, and the minimum of the LJ potential is located at the vertical position of the back scatterer row (results for a lattice wall model consisting of four scatterer rows appear in SI Section 4, showing qualitatively similar results as those of the two-row wall model). We chose an inter-scatterer distance of $d = 3$ Å and considered initial Gaussian wavepackets of widths $\sigma_x = \sigma_y = 1, 3$, and 5 Å, carrying angular momentum of $m = +1$ and given center of charge velocity of $v_y = 30$ Å/fs towards the wall (more simulation parameters are given in the caption of Fig. 6).

When the ratio between the initial wavepacket width and the interlayer spacing is smaller than 1 $(\tau \equiv \sigma_x/d = \sigma_y/d < 1)$, the wavepacket interacts locally with its adjacent scattering sites and splits (Fig. 6b, top panels and Supplementary Movie 4). The two sub-packets scatter mostly backwards, with some tendency to the right for a head-on collision with one of the scattering sites (top right subpanel of Fig. 6b) or to the left for a bond collision in between two scattering sites (see top middle subpanel of SI Fig. S5). The overall deflection (blue lines in Fig. 6c) is an order of magnitude weaker than that observed for the single-scatterer diffraction case (Fig. 5c), as may be expected for a less corrugated surface. For $\tau = 1$, both head-on and bond collisions result in clear tendency to the right (Fig. 6b, middle panels), but the reflected wavepacket remains more localized hence the overall sideways deflection is even smaller (orange lines in Fig. 6c). Finally, when $\tau = 5/3$ the collision is delocalized along the scattering surface (Fig. 6b, bottom middle panel), and the reflected wavepacket is much more symmetric (Fig. 6b, bottom right panel) with further reduced sideways deflection, regardless of the center of charge collision site (green lines in Fig. 6c). Notably, angular momentum inversion is clearly observed in all deflecting wavepackets (Fig. 6d and Supplementary Movie 5). Moreover, the $m = 0$ wavepacket displays fully symmetric backscattering, whereas the $m = -1$ wavepacket shows the same behavior as its $m = +1$ counterpart but with mirror image symmetry around the central vertical axis (see Fig. S8 in SI Section 5). This demonstrates OAM-based spatial resolution induced by the surface scattering process.

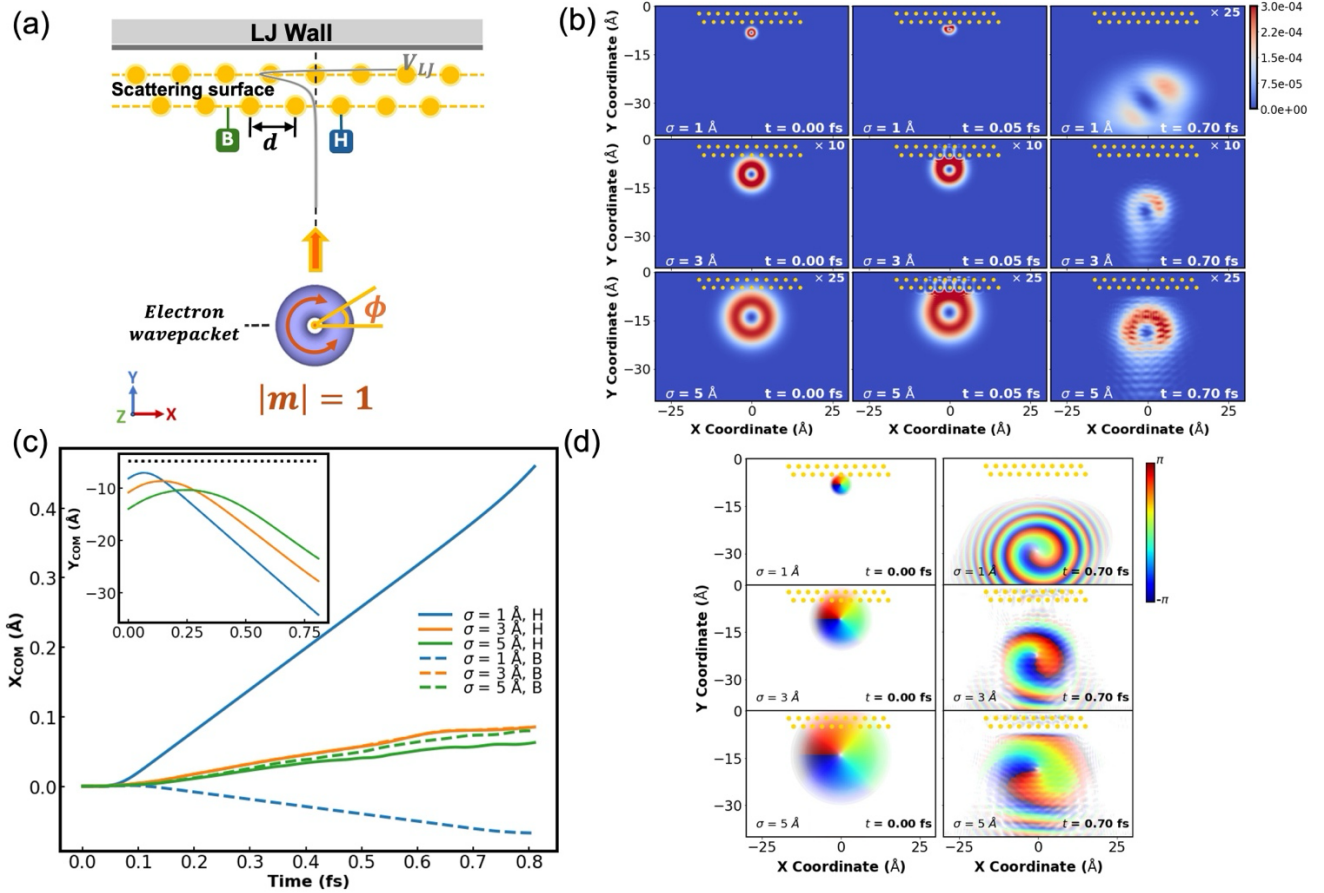

**Figure 6.** (a) Schematic illustration of the scattering of a 2D Gaussian electronic wavepacket, carrying angular momentum of $|m| = 1$, from a two-row fixed scatterer surface. The scatterer rows are laterally shifted with respect to each other by half a lattice constant, $0.5d = 1.5$ Å. A vertical LJ potential with $\varepsilon_{LJ} = 0.05$ eV is introduced to mimic scattering from bulk layers, with its minimum located at the inner scatterer row ($y_{min} = -2.25$ Å, grey curve). The center of charge of the wavepacket is initially located either in front of one of the first-row scatterers (head-on collision, marked as H) or in between two such scatterers (bond collision, marked as B). (b) Snapshots taken from head-on collision simulations of wavepackets of initial widths $\sigma = 1$ (top row), 3 (middle row), or 5 Å (bottom row), positioned $-8.2$, $-10.9$, or $-14.0$ Å below the scatterers line, respectively, and given OAM of $m = +1$ and a vertical velocity of $v_y = 30$ Å/fs towards the surface. The snapshots are taken at $t = 0$ (left column), 0.05 (middle column), and 0.7 fs (right column). (c) Time evolution of the transversal coordinate expectation value, $\langle x \rangle$, of the wavepackets presented in panel (b). Head-on and bond collision results are presented by the full and dashed lines, respectively. All other simulation parameters are the same as in Fig. 5(b). The inset presents the time evolution of the vertical coordinate expectation value, $\langle y \rangle$, where the black dashed line represents the position of the first scatterer row. (d) Phase-resolved scattering wavepacket snapshots extracted at $t = 0$ and 0.7 fs.

Enhancement of the OAM spatial resolution can be achieved by increasing the corrugation of the scattering surface. To demonstrate this, we introduce an impurity scatterer atop the surface in front of the wavepacket (see Fig. 7a). Similar to the case of an isolated obstacle, when carrying finite OAM the wavepacket demonstrates strongly asymmetric scattering, but rather than diffracting it is now reflected backwards due to the presence of the underlying wall. The strongest asymmetry is obtained for the most localized wavepacket ($\sigma = 1$ Å, top panels of Fig. 7a and blue lines in Fig. 7b) with center of charge sideways deflection comparable to that of the single-scatterer case (Fig. 5c), regardless of the position of the surface impurity (full and dashed lines in Fig. 7b). As the initial width of the wavepacket increases significant splitting of the scattered wavepacket is obtained (middle and bottom panels of Fig. 7a) and the backscattering asymmetry reduces (orange and green lines in Fig. 7b) but remains considerably larger than that observed for the flat scatterer surface model. Mirror image results are obtained for wavepackets carrying opposite initial angular momentum, and symmetric scattering is obtained for $m = 0$ wavepackets (see SI Section 5).

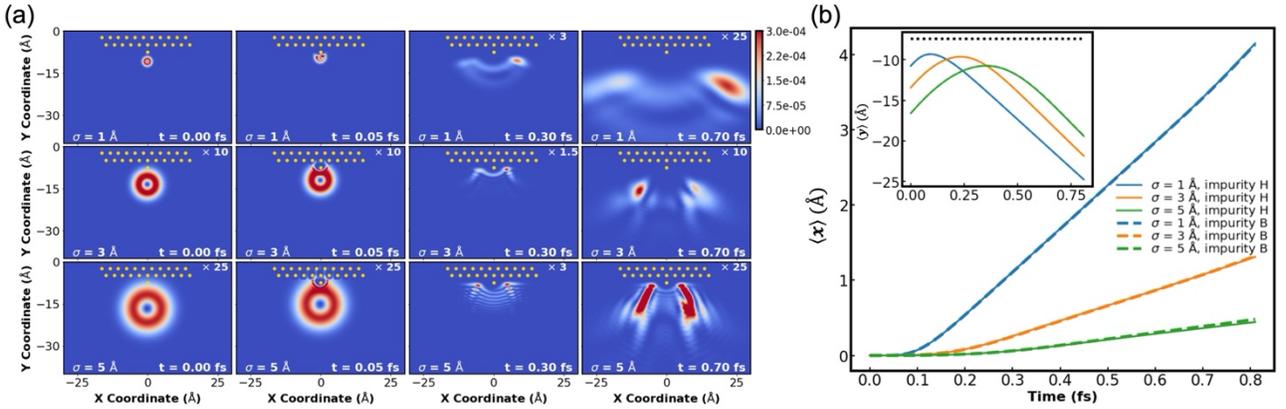

**Figure 7.** (a) Snapshots taken from head-on collision simulations of 2D Gaussian electronic wavepackets of initial widths $\sigma = 1$ (top row), 3 (middle row), or 5 Å (bottom row), positioned $-10.78$, -13.45, and -16.6 Å, respectively, directly below an impurity scatterer (placed in front of a surface site, marked as H position) and given OAM of $m = +1$ and a vertical velocity of $v_y = 30$ Å/fs towards the corrugated surface. The snapshots are taken at $t = 0$ (first column), 0.05 (second column), 0.3 (third column) and 0.7 fs (fourth column). (b) Time evolution of the transversal coordinate expectation value, $\langle x \rangle$, of the wavepackets presented in panel (a). Blue, orange, and green lines represent results for initial wavepacket widths of $\sigma = 1, 3,$ and 5 Å, respectively. Results for H-positioned impurities (full lines) and B-positioned impurities (dashed lines), where the impurity is positioned at a bond site, are presented. The inset presents the time evolution of the vertical coordinate expectation value, $\langle y \rangle$, where the black dashed line represents the vertical position of the impurity scatterer. All other simulation parameters are the same as in Fig. 6.

The abilities demonstrated above, to induce well-defined electronic OAM and spatially resolve it, are key ingredients in the emerging field of orbitronics, where orbital angular momentum (in addition to electronic charge and spin degrees of freedom) carries and conveys quantum information[57-60]. OAM polarization is achieved by forcing the electrons to traverse a chiral molecular or solid medium, whereas OAM-based spatial resolution involves scattering from an atomically corrugated surface. The introduction of spin-orbit coupling, e.g., from the scattering surface, may associate the rotational sense of the wavepacket with a specific spin orientation. Hence, spatial resolution based on the OAM accumulated by electrons traversing a chiral medium may be manifested as chirality-induced spin selectivity, like that observed in photoemission experiments through chiral molecular layers followed by Mott polarimetry. Future efforts to further enhance OAM-based spatial resolution may rely on the introduction of electronic friction at the scattering surface (see SI Section 2).

## Methods

For simulating OAM polarization of electronic wavepackets traversing a chiral medium we constructed a three-dimensional simulation box of dimensions ranging from $(-15, -15, -5)$ Å to $(15,15,45)$ Å, in which the wave function is represented on a Cartesian grid of uniform spacing of $0.1$ Å. The chiral medium was represented by a fixed left-handed helix of the following parametric equation:

$$\boldsymbol{h}(\theta) = (R_h \cos(\theta), -R_h \cos(\theta), P_h \theta/2\pi), \quad [7]$$

where $\theta$ is the azimuthal angle, $P_h$ denotes the helix pitch, and $R_h$ represents the helix radius. Discrete positive point charges were uniformly distributed along the helix (see Fig. 1a). The potential exerted by each point charge was taken to be of Coulombic attraction nature, $V_c(x, y, z) = -\frac{q_c}{4\pi\varepsilon_0 \rho}$, where $\varepsilon_0$ is the vacuum permittivity, $q_c$ is the (positive) particle charge, and $\rho = [(x - x_c)^2 + (y - y_c)^2 + (z - z_c)^2 + R_\eta^2]^{1/2}$, where $\boldsymbol{r}_c = (x_c, y_c, z_c)$ is the position of the point charge and $R_\eta = 0.01$ Å is a smoothening parameter introduced to avoid singularities. The electronic Gaussian wavepacket (see Eq. [1]), which was initially placed with its center of charge positioned at the origin, was driven through the helix by an external static vertical electric field of $\boldsymbol{E} = (0,0, -5)$ V/Å. The quantum wavepacket dynamics was described using the time-dependent Schrödinger equation propagated using the 4th order Runge-Kutta scheme, with circular boundary conditions applied at the box boundaries along the three

Cartesian axes. A step-doubling algorithm was employed with an initial time increment of 0.1 attosecond. Here, one constantly compares the total energy and normalization obtained after two consecutive time-steps to those obtained after a single double-step (with double the time increment). If the relative energy difference and the normalization difference are both lower than $10^{-4}$ then the two-steps wavefunction is adopted, the time increment is up-scaled by a factor of 1.01, and the propagation proceeds. Otherwise, the wavefunction is rolled-back to the previous time-step, the time increment is down-scaled by a factor of 0.99, and the propagation repeats. The expectation value of the OAM operator component along the main axis of the helix (the z direction) was determined by $\langle \hat{L}_z \rangle = \int \psi^*(x,y,z) \hat{L}_z \psi(x,y,z) dx dy dz$, where $\hat{L}_z = -i\hbar \left( x \frac{\partial}{\partial y} - y \frac{\partial}{\partial x} \right)$, with $\hbar$ being the reduced Planck constant [54], and the spatial numerical derivatives are evaluated using a 9-point stencil.

The scattering simulations involved 2D Gaussian electronic wavepackets (see Eq. [2]) scattered off different surface models: (i) a smooth LJ wall (see Fig. 4); (ii) a single scatterer (see Fig. 5); (iii) a corrugated two-layer scatterer wall augmented by a LJ potential (see Fig. 6); and (iv) a corrugated two-layer scatterer wall augmented by a LJ potential and an impurity scatterer (see Fig. 7). Two-dimensional simulation boxes were constructed with dimensions ranging from $(-60, -70)$ Å to $(60, 0)$ Å, for the LJ wall simulations; from $(-30, -80)$ Å to $(30, 0)$ Å, for the single scatterer simulations; and from $(-30, -50)$ Å to $(30, 0)$ Å, for the corrugated wall simulations. The wave function was represented on a Cartesian grid of uniform spacing of 0.05 Å. Sensitivity tests of the scattering results towards the choice of scattering potential parameters are presented in SI Section 6.

## Data availability

All data supporting the findings of this study are available within the article, the Supplementary Information file, and the Supplementary Movie files. Additional raw data and details of the analysis procedures are available from the corresponding authors upon request.

## Acknowledgments

O.H. is grateful for the generous financial support of the United States – Isarel Binational Science Foundation via NSF-BSF grant number 2023602, the Israel Science Foundation via grants numbers 3645/24 and 3646/24, and the Heinemann Chair in Physical Chemistry. A.N.'s work was supported by the National Science Foundation Grant # 2451953.

## Author contributions

O. H., J. G. and A. N. conceived the research idea, and all authors contributed to designing the study. Y. C. performed the simulations and collected the data. All authors contributed to data analysis, discussion, and manuscript preparation.

## Competing interests

The authors declare no competing interests.